\documentclass[aps, prb, reprint, amsfonts, amssymb, amsmath]{revtex4-1}
\usepackage{graphicx}

\begin{document}

\title{Atomic discreteness and the nature of structural equilibrium in conductance histograms of electromigrated Cu-nanocontacts}
\date{\today}

\author{Robert \surname{Pfender-Siedle}$^{1,2}$, Julia \surname{Hauser}$^{1}$, and Regina \surname{Hoffmann-Vogel}$^{1,2}$}
\affiliation{$^{1}$Physikalisches Institut, Karlsruher Institut f\"ur Technologie, D-76128 Karlsruhe, Germany\\
$^{2}$Institut f\"ur Angewandte Physik, Karlsruher Institut f\"ur Technologie, D-76128 Karlsruhe, Germany}

\begin{abstract}
We investigate the histograms of conductance values obtained during controlled electromigration thinning of Cu thin films. We focus on the question whether the most frequently observed conductance values, apparent as peaks in conductance histograms, can be attributed to the atomic structure of the wire. To this end we calculate the Fourier transform of the conductance histograms. We find all the frequencies matching the highly symmetric crystallographic directions of fcc-Cu. In addition, there are other frequencies explainable by oxidation and possibly formation of hcp-Cu. With these structures we can explain all peaks occurring in the Fourier transform within the relevant range. The results remain the same if only a third of the samples are included. By comparing our results to the ones available in the literature on work-hardened nanowires we find indications that even at low temperatures of the environment, metallic nanocontacts could show enhanced electromigration at low current densities due to defects enhancing electron scattering.
\end{abstract}

\maketitle
\newpage

\section{Introduction}

Since the functional elements in electronic devices continuously become smaller it becomes increasingly important to understand the properties of small metallic leads to these devices at the nanometer scale. The metallic leads are currently made out of Cu, therefore the electronic transport properties and the electromigration properties of Cu nanocontacts are of particular technological importance. Lately, the interplay of structural equilibrium and the conductance has been debated, when work-hardened wires have been compared to annealed ones and even heated and electromigrated wires \cite{Yanson2005,Hoffmann2008}. The conductance of nanocontacts of different metals has been investigated mainly using mechanically controlled break junctions \cite{Muller1992, Hansen1997, Rodrigues2000, Bakker2002, Krans1995}. With this technique a free-standing metallic bridge is fabricated on a flexible substrate, and by bending the substrate, the metallic bridge is elongated and thinned until first a nanocontact is formed and finally until breakage \cite{vanRuitenbeek1996}. The nanocontacts show stepwise conductance changes on the order of the conductance quantum $G_0=2e^2/h$. The conductance quantum is the conductance of a single electronic channel of transparency 1. By using data of more breaking processes and plotting it in conductance histograms the quantum nature of electronic transport in nanocontacts was confirmed \cite{Olesen1995, Krans1995, Gai1996}.

On the other hand, also the atomic structure leads to conductance jumps since the diameter of the contact must change discontinuously due to the atomistic nature of matter. Yanson {\it et al.} investigated work-hardened Au-break-junction-nanocontacts and also found stepwise conductance changes \cite{Yanson2005}. By analyzing these further they found additional periodic structures that can be explained by the discreteness of the contact size for the three highly symmetrical directions of the Au fcc lattice: The conductance of a nanocontact is in direct relation with the cross section \cite{Torres1994}. This was also confirmed experimentally by TEM analysis of $(110)$-oriented Au wires \cite{Kurui2009}. The same characteristic conductance peaks were also found for electromigration of gold \cite{Hoffmann2008}.
Other metals also show different conductance peaks when work-hardened. Aluminum \cite{Yanson2008}, copper and silver \cite{Shklyarevskii2013} all show more peaks when work-hardened compared to annealed. These metals also show a characteristic periodicity, but it only corresponds to one direction of the fcc-lattice, $\left(111\right)$ in the case of aluminum and $\left(100\right)$ for copper and silver.

Dreher et al. have calculated the structure and conductance of Au nanocontacts and found a periodicity in the minimal radius of the cross-section that was related to the area occupied by one atom \cite{Dreher2005}. However, in their calculation several configurations contributed to each conductance peak they observed such that the periodicity did not show in the conductance histogram. The histogram generated from their data appears closely related to the histograms observed for annealed contacts by the break junction technique rather than the ones observed for electromigrated or work-hardened nanocontacts. This could be due to conditions such as temperature or annealing time that match the experimental conditions of annealed nanocontacts better than the ones for electromigrated or work-hardened nanocontacts.

In this work we investigate conductance histograms of electromigrated Cu-contacts. We find periodic structures that can be explained by the three highly symmetric directions of the lattice. We suggest oxidation and formation of a hcp-lattice as causes for additional frequencies. The frequencies related to these structures explain all peaks in the Fourier transform of the conductance histogram in the relevant frequency range. We investigate the relative peak height and find a similar relative peak height for pure and oxidized Cu for the $\left(100\right)$-direction with respect to the $\left(110\right)$-direction. Part of the data shows all peaks observed in the full data set. Hcp-Cu peaks are absent for larger values of the conductance. We compare our results to literature and discuss how additional defects in the metal could lead to enhanced electron scattering and local heating enhancing electromigration even at low current densities and low temperatures of the environment.

\section{Experimental methods}

Cu nanostructured thin films were prepared by sputtering thin layers of copper onto a silicon-substrate covered by a 500 to 1000-nm-thick thermally oxidized layer for electrical insulation. The pressure of the vacuum chamber was below $10^{-3}$ Pa. To define the geometry of the films, during sputtering stainless steel masks with slit-shape openings were placed in front of the substrate. The length of the slit was between 4 and 8 mm and its width between 30 and 500 $\mu$m. The sputtering rate was adjusted to $18$ nm/min. The film thickness varied between 27 and 54 nm. Part of the samples were contacted using silver paste, others were contacted using Al bond wire.

The nanostructured films were then subject to controlled electromigration in air and at room temperature, using an algorithm first proposed by Strachan et al. \cite{Strachan2005} slightly modified as described in the following. The main idea is to avoid thermal runaway often occurring during electromigration at constant voltage by using the temperature dependence of the resistance to control the temperature at the forming contact. During an electromigration cycle the voltage was increased in a stepwise manner until heating was detected as an absolute increase of the resistance. For a pre-chosen absolute increase of the resistance, the cycle was stopped. The next cycle was then started with 60\% of the last voltage to allow the contact to cool down. The resistance value for stopping the cycle was increased by a pre-chosen step.

The advantage of an absolute resistance value exit condition compared to a relative exit condition as used previously \cite{Hoffmann2008,Stoeffler2014} is that it allows a smooth increase of the resistance even for several orders of magnitude. At the start of the thinning process, a relatively large increase of the resistance is needed to start irreversible changes of the resistance as compared to reversible resistance changes due to heating. At later stages, such large resistance steps could lead to overheating and destruction of the thin film. To achieve similarly good results with a relative exit condition \cite{Hoffmann2008}, the relative changes of the resistance had to be adjusted several times at the start of the electromigration process, making the process dependent on human interaction.

\begin{figure}[b]
	\includegraphics[width=1.0\linewidth]{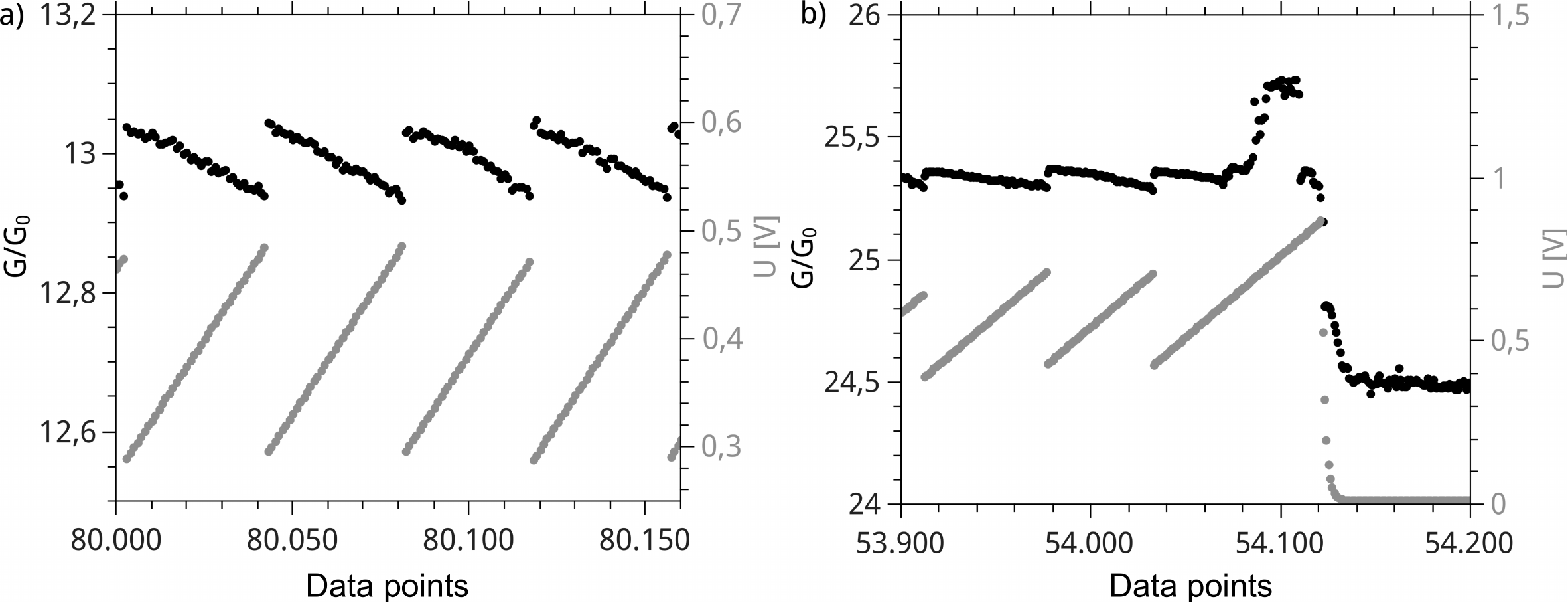}
	\caption{\label{fig-1} Evolution of conductance as a function of applied voltage during controlled electromigration thinning.}
\end{figure}

During the full electromigration thinning process the current and imposed voltage were measured with a Keithley 2400 sourcemeter. To obtain the conductance the current was divided through the voltage and the starting conductance was subtracted. How the voltage evolves during typical electromigration cycles as a function of data point is shown in Fig. \ref{fig-1}. Since for each data point, a fixed time is allowed, the $x$-axis translates into a time axis. The conductance clearly depends on voltage, in a mostly linear way. Nonlinearities have been investigated for Au contacts in \cite{Stoeffler2014} where we found nonlinearities occcurring in parallel to metallic conduction due to tunneling contacts. These nonlinearities were found to be small in the data range investigated here. Finally the conductance data was plotted in a histogram as discussed below.

Several points should be noted when the data are compared to data resulting from other thinning methods: 1. The goal of the thinning process is the increase of the resistance. Therefore the voltage is increased within one cycle. As a result, resistance values resulting from a range of voltages are included in the same histogram. At the end of each cycle, one particular atomic configuration is frozen in because the contact is cooled down. This atomic configuration is counted particularly often, because the contact remains in this atomic configuration. Other possible energetically favorable configurations may not be reached during one particular cycle and could as a result obtain less counting statistics in a histogram. 2. Due to heating resulting from the dissipated electrical power, the contact resistance is increased as a function of voltage and each peak is broadened. The main dependence on voltage is a linear one. Nonlinear dependencies attributed to tunneling contacts in parallel to metallic conduction have been investigated for Au in \cite{Stoeffler2014}. 3. The length of each plateau of constant resistance is not only given by thermal activation and by the stability of the atomic configuration but also by the step size of the resistance increase exit condition. This length is related to the counting numbers of the resistance value in the histogram. This property of the algorithm enhances counting statistics for small conductance values, because $R \sim G^{-1}$ and for large jumps in $R$. The counting statistics is therefore not a measure of the probability of an atomic configuration to occur, but also of the step size in $R$ needed to reach it.

\section{Results and discussion}

\begin{figure}[b]
	\includegraphics[width=0.8\linewidth]{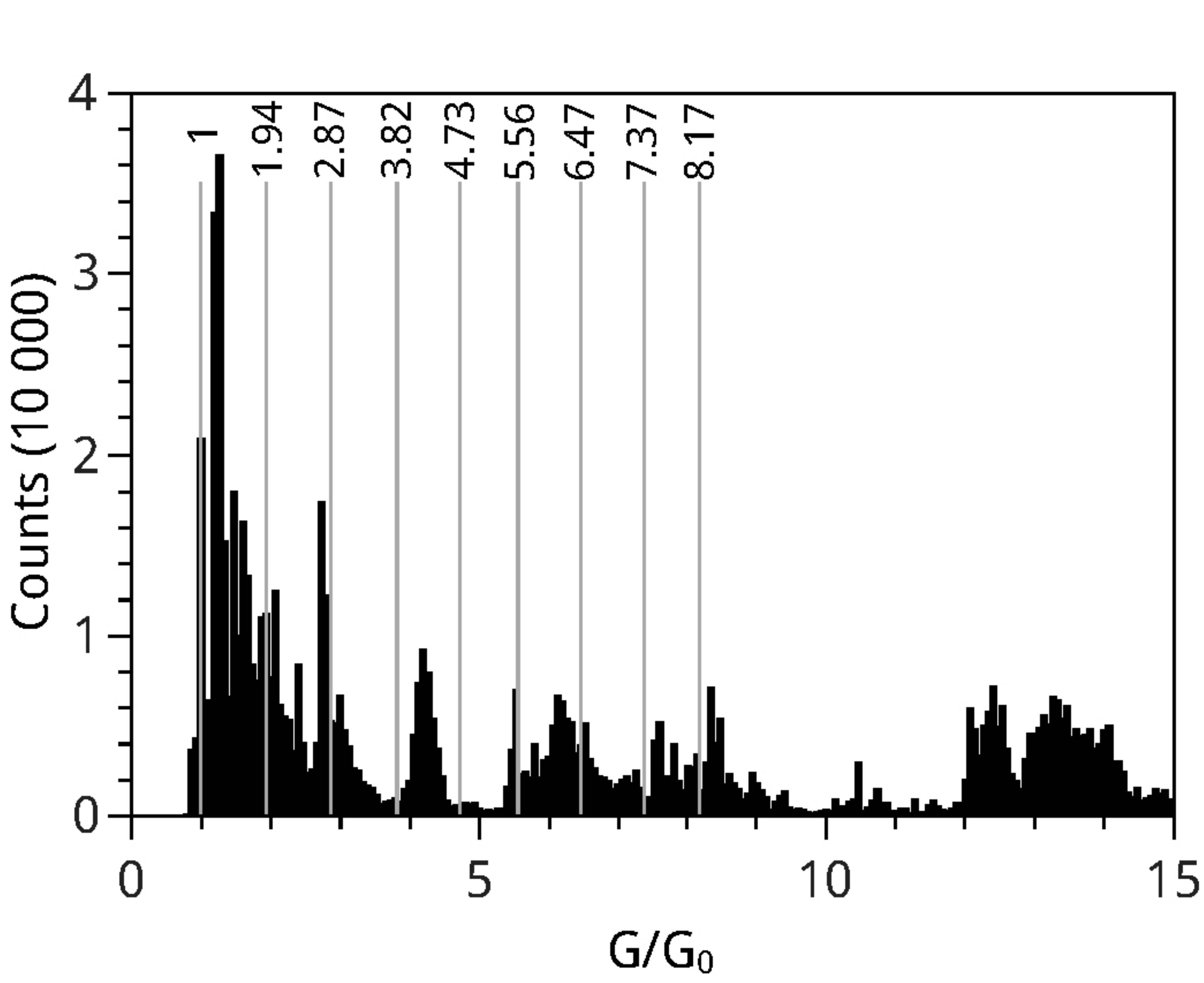}
	\caption{\label{fig-2} Conductance histogram of electromigrated copper nanocontacts (black bars). The conductance peaks of work-hardened mechanically controllable copper break junctions \cite{Shklyarevskii2013} are marked with red lines. These peaks result from the crystallographic $\left[100\right]$-direction of Cu being perpendicular to the cross-section of the nanocontact.}
\end{figure}

We used data from 40 Cu-nanocontacts to generate the conductance histogram shown in Fig. \ref{fig-2}. 33 of them showed data in the desired conductance range. When comparing the histogram to the one of copper break junctions \cite{Shklyarevskii2013}, we find most of the peaks found earlier. These peaks were identified previously as resulting from the crystallographic $\left[100\right]$-direction being perpendicular to the cross-section of the nanocontact \cite{Yanson2008}. Since in the plane of the cross-section, the diameter cannot be varied continuously, but only atom by atom, the area of the cross-section varies stepwise, leading to peaks in the conductance histogram. The conductance of a wire with infinitely high potential walls at its outer limits has been calculated \cite{Torres1994}

\begin{equation}
 G\approx G_0\left(\pi A-\frac{L}{2}\right)
 \label{eqn-leitwert}
\end{equation}

where $A$ is the cross-section and $L$ is the circumference of the wire both in units of the Fermi wavelength $\lambda_F$. For finite walls, there is a leakage of the electronic wave-functions into the environment, increasing the radius of the wire from $R$ to $R'=R+0.17\cdot\lambda_F$ for a description in good approximation \cite{Ruitenbeek1997,Yanson2005}. Inserting this value to equation \ref{eqn-leitwert}, we arrive again at the well-known Sharvin formula \cite{sharvin65}, linking the conductance of a short cylindric nanocontact to its cross-section:

\begin{equation}
 g=\frac{G}{G_0}\approx \pi A
 \label{eqn-Sharvin}
\end{equation}

In this view the conductance change caused by the addition of one atom mainly results from the area added to the cross-section of the contact assuming one channel per atom for the monovalent Cu. For the crystallographic $\left(100\right)$-direction as the area of the contact, one obtains $\Delta g_{\text{fcc-Cu,}(100)}=0.958$ using $a=3.62$ \AA~and $k_F=1.36/$\AA~ for Cu \cite{ashcroft,Yanson2008}. In the histogram generated from measured data (black bars in Fig. \ref{fig-2}), only the fourth and fifth predicted peak (red lines) correspond to small maxima, while all other predicted peaks correspond to clear maxima.

Additional peaks observed in our histogram, notably for example at 1.2; 4.2 and 6.3 $G_0$, are not present for copper break junctions. For Au nanocontacts, it had been found that electromigrated nanocontacts showed more peaks than annealed contacts studied with the mechanically controllable break-junctions \cite{Hoffmann2008}. The Au peaks were explained by the $\left(100\right)$ direction as well as by other crystallographic directions contributing to the cross-section of the atomic contacts, the $\left(111\right)$ and the $\left(110\right)$ direction \cite{Yanson2005}. These are the most stable low-indexed crystallographic directions for fcc metals. The periodicities corresponding to these directions were identified in the Fourier transform of the histogram for Au, see Ref. [\onlinecite{Yanson2005}]. The presence of additional peaks for the electromigrated Cu contacts suggests that for contacts thinned by electromigration there are additional stable atomic configurations compared to contacts studied using the mechanically controllable break-junction method.

\begin{figure}[t]
	\includegraphics[width=0.9\linewidth]{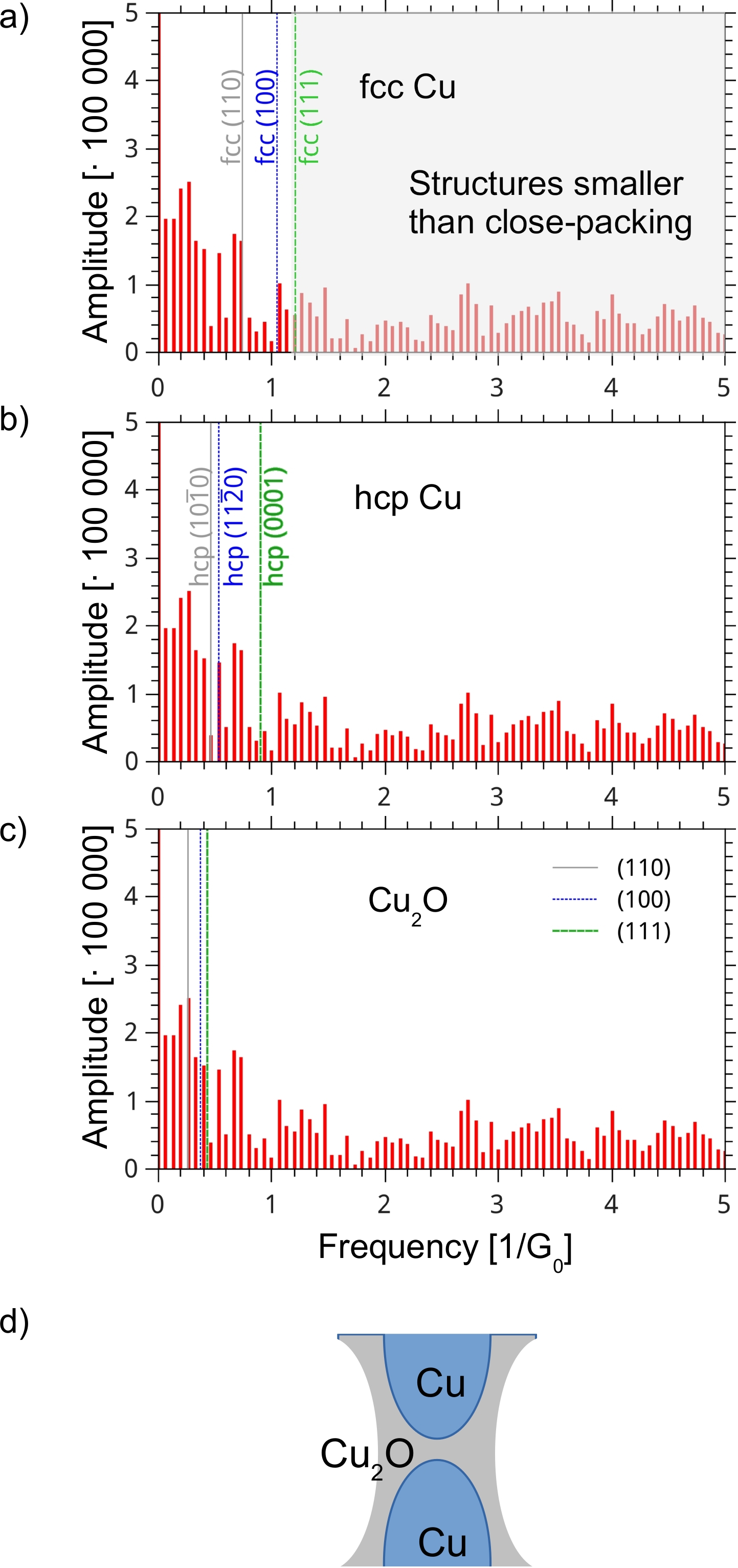}
	\caption{\label{fig-3} Fourier transform of the conductance histogram with the calculated frequencies for the three highly symmetric directions in the (a) fcc-lattice and (b) hcp-lattice. (c) Highly symmetric directions for Cu(1)oxide assuming the Fermi wavelength of Cu. (d) Model of the Cu contact surrounded by Cu(1)oxide.}
\end{figure}

To check whether the additional peaks in the conductance histogram arise from the other low-indexed crystallographic fcc directions, we calculated the Fourier transform of the conductance histogram to see frequencies and periodicities in the conductance data (Fig. \ref{fig-3}). We use the same bin-size and data range (0 to 15 $G_0$) as Ref. [\onlinecite{Yanson2005}]. Indeed, also in the Fourier transform, we observe a number of peaks. The one at $f=\frac{1}{\Delta g}$ with $\Delta g_{\text{fcc-Cu,}(100)}=0.958$ inferred from the conductance histogram is indeed found. The experimentally obtained peak at $\Delta g=0.93$ occurs at slightly smaller values compared to the predicted peak calculated using equation \ref{eqn-Sharvin}, by about half a bin size, $(1/15) (1/G_0)$. A similar deviation had been observed before in the case of Au, see Ref. [\onlinecite{Yanson2005}]. For the additional expected periodicities for the fcc lattice, we obtain $\Delta g_{\text{fcc-Cu,}(111)}=0.830$, and $\Delta g_{\text{fcc-Cu,}(110)}=1.355$ with $f=\frac{1}{\Delta g}$ using equation \ref{eqn-Sharvin}. Figure \ref{fig-3}(a) shows that the frequencies we expect are found with only small deviations namely the frequency calculated for the $\left(111\right)$ direction is slightly smaller than the one of a nearby measured peak at $\Delta g=0.79$.

The Fourier transform data shows a number of peaks in addition to the three identified above. This suggests that additional structures could occur in Cu-nanocontacts. The main relevant area where the peaks could be explained by the method detailed above is the one below $(\Delta g_{(111)})^{-1}$, because large frequencies correspond to small atomic sizes, and in the area above $(\Delta g_{(111)})^{-1}$, it would be necessary to change the area of the nanocontact by fractions of Cu atoms, since the $\left(111\right)$ direction is the closest packed possible arrangement of Cu.

For a number of metals it is well-known that atomic structures differing from the bulk ones are possible in thin layers or nanostructures, for example Fe, usually present in the bcc phase, is known to form a fcc phase as a thin film. As another example, Au is known to show a hcp and an fcc phase within its herringbone reconstruction at clean surfaces. For Cu, in addition to the standard fcc phase, mainly the existence of hcp Cu is discussed in literature. Jona et al. \cite{Jona2003} reported the growth of a $\left(11\bar{2}0\right)$ hcp-Cu-film on a W $\left(001\right)$ surface. Calculations showed two energy minima for hcp-Cu states neither of which is stable in bulk. Using $a=2.96$~\AA~ and $c=5.01$~\AA~ (taken from \cite{Jona2003}) we calculated the expected periodicities for a hcp-Cu-lattice, see appendix. We obtain $\Delta g_{\text{hcp-Cu,}(10\bar{1}0)}=2.180$, $\Delta g_{\text{hcp-Cu,}(0001)}=1.114$ and $\Delta g_{\text{hcp-Cu,}(11\bar{2}0)}=1.888$ and compare to the experimental data in Fig. \ref{fig-3}(b). The frequency for $\left(11\bar{2}0\right)$ matches an experimentally found one, while the other two directions of the hcp-lattice are not observed.

To explain additional frequencies we suggest oxidation. We have observed the oxidation of macroscopic Cu wires under the influence of controlled electromigration using scanning electron microscopy with elemental analysis on a cross-section of the wires \cite{Hauser2015}. The oxide layer at the surface of the wire mechanically stabilizes macroscopic Cu wires while the inner part of the wire remains unoxidized. At room pressure the thin films studied here are subject to a large dose of oxygen. Cu does not passivate under the influence of oxygen, and as a consequence we expect full oxidation of the initially clean Cu layer as a function of time. On the other hand Cu(1)- and Cu(2)-oxides are insulators, and since we perform electromigration experiments and observe metallic conduction characteristics of the thin film, there must be a part of the thin film remaining unoxidized at the time when the measurement was performed. We therefore take the assumption that only part of the thin film, possibly a surface layer, is oxidized. We assume that the oxide in the vicinity of the conductive areas has Cu(1)-oxide properties since the oxide results from oxygen moving from the surface towards the center of the thin film as a function of time with the oxygen content gradually increasing. Cu(1)-oxide has a cubic structure where the Cu atoms assume the fcc structure and the oxygen atoms assume a bcc structure. The lattice constant is $a=4.27$ \AA, see Ref. \cite{Wells1984}. Since the electronic transport remains metallic, we assume that the electrons still assume the Cu Fermi wavelength given above.

Using these assumptions we obtain $\Delta g_{\text{Cu$_2$O,}(100)}=2.68$ and $\Delta g_{\text{Cu$_2$O,}(110)}=3.80$ and $\Delta g_{\text{Cu$_2$O,}(111)}=2.32$. Comparing these predictions to the experimentally obtained values (Fig. \ref{fig-3}c)), again the deviations are smaller than the bin size we use.

The relative peak height is influenced by the electromigration algorithm and by the individual history of the thinning process for each sample. In this view it is surprising that for the pure Cu fcc peaks and for the Cu(1)-oxide peaks, the height of the peak at $\Delta g_{(100)}$ is about $0.6$ times the height of the peak found at $\Delta g_{(110)}$. This could be explained if the general relative abundance of $\left(100\right)$-oriented grains with respect to $\left(110\right)$-oriented grains remains constant throughout the oxidation process.

\begin{figure}[t]
	\includegraphics[width=0.7\linewidth]{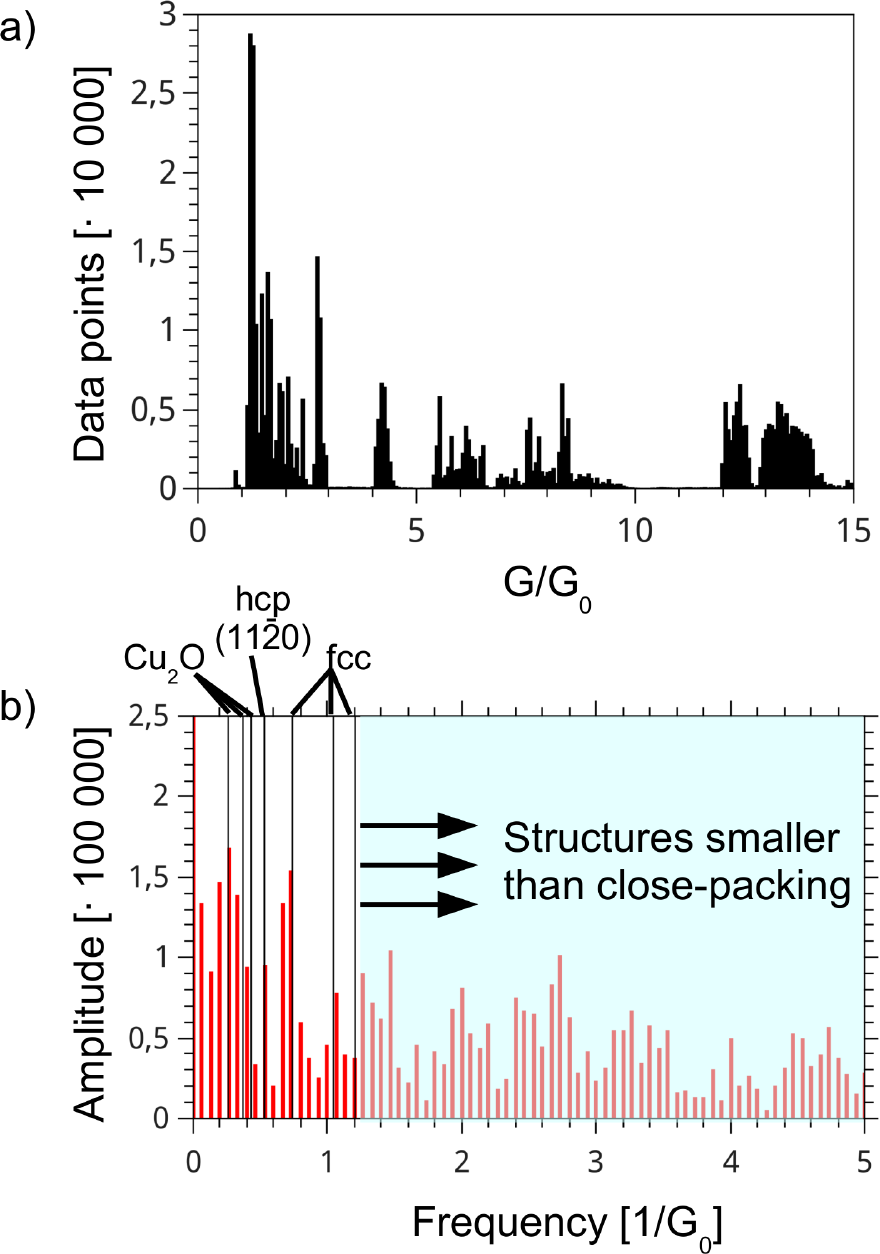}
	\caption{\label{fig-4} a) Conductance data and b) Fourier-transform if the conductance data for a third of the samples. In b) all the peaks identified from the full data set are included as black lines.}
\end{figure}

It should be carefully discussed whether the observed peaks really result from atomic periodicities or are just due to noise. One way of checking the reproducibility of our results is to analyze only part of our data. Using the conductance data of only 10 nanocontacts, obviously we obtain a smaller total amount of data points, see Fig. \ref{fig-4}. In addition, we obtain conductance regions where no values have been observed by chance as a result of the natural thermally activated fluctuations of the atomic positions during controlled electromigration. The Fourier transformed conductance histogram shows all the peaks that have been identified before, also with a similar deviation between the calculated frequency and the measured one for the $\left[111\right]$ direction.

\begin{figure}[t]
	\includegraphics[width=0.7\linewidth]{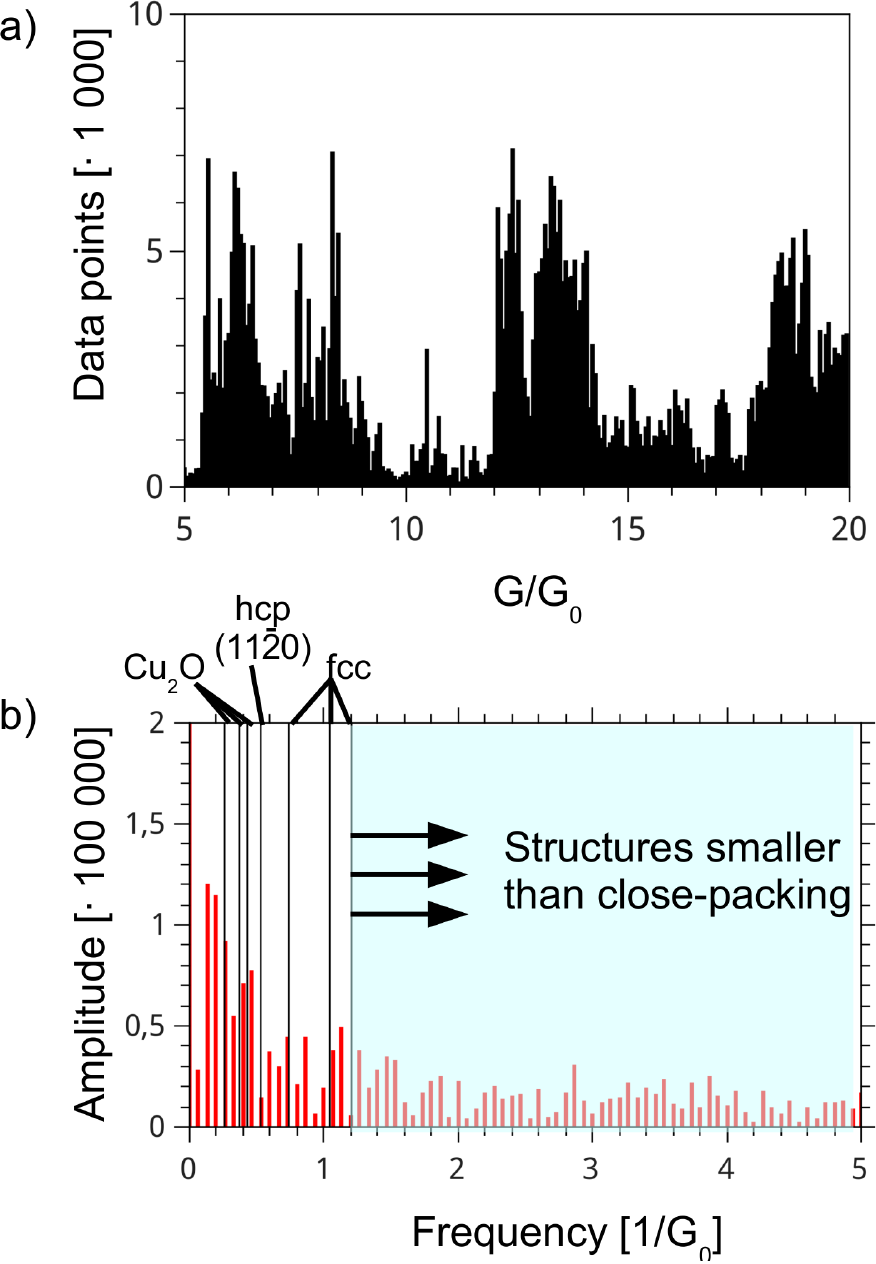}
	\caption{\label{fig-5} a) Conductance data and b) Fourier-transform if the conductance data for conductance values between 5 and 20 $G_0$. In b) all the peaks identified from the full data set are included as black lines.}
\end{figure}

Another validity check is obtained when we analyze the conductance range between $5$ and $20 \,G_0$ moving the window of analysis by $5 \,G_0$ to larger values, see Fig. \ref{fig-5}. Again, the number of data points is smaller in this higher conductance regime compared to the data set at low conductance values.
For this data range, we expect the Cu nanocontacts to be larger in size. Again we show the original data Fig.~\ref{fig-5}a) and the Fourier-transform Fig.~\ref{fig-5}b). In the latter graph we have included again all peaks identified from the full data set marked as black lines. In this Fourier-transform, additional unidentified peaks occur near  0.14 $G_0^{-1}$ and 0.87 $G_0^{-1}$. The peak near 0.87 $G_0^{-1}$ occurs close to the peak expected for the hcp $\left( 0001 \right)$-direction, but since this peak hadn't been observed neither in the full data set nor in the data generated from 10 samples, it seems improbable that this should be a good explanation for this peak. Instead the peaks at 0.14 $G_0^{-1}$ and 0.87 $G_0^{-1}$ could be explained by the formation of a strained layer in the contact, where the lattice constant is locally increased. Such a strained layer could be caused by repeated heating and thermal expansion. The data shows no peak at the position where the hcp $\left( 11\overline{2}0 \right)$ peak is expected in contrast to the full data set and to the partial data set from 10 samples shown above. Since the hcp structure is stable only for a few number of atomic layers in thin films, this could be a sign that this structure is no longer stable for larger nanocontacts as might be expected.

For Au nanocontacts, it had been found that electromigrated nanocontacts showed more peaks than annealed contacts studied with the mechanically controllable break-junctions but similar peaks as work-hardened contacts studied with this method \cite{Hoffmann2008}. It is surprising that the number of peaks is large for both work-hardened Cu and electromigrated Cu nanocontacts, but not for mechanically stretched nanocontacts from annealed material. Work-hardening enhances the number of defects such as dislocations in a metal, whereas annealing, as occurring during electromigration, decreases the number of defects - this seems contradictory. For both Au and Al the conclusion was drawn that work-hardened wires had additional degrees of freedom that allowed for more atomic configurations close to equilibrium \cite{Yanson2005, Yanson2008}. However, this understanding of mechanical equilibrium seems to contradict the common understanding that defects disturb structural equilibrium rather than enhancing it. In this picture it does not become clear why defects should mechanically enhance equilibrium. Another way of understanding these experimental facts may be that the presence of a more defective microstructure of the nanocontact could cause locally enhanced scattering leading to a locally enhanced hopping probability triggered by electromigration that occurs even at small current densities and low temperatures of the environment of the metallic structure.

Analyzing the data as a function of $\sqrt{g}$ shows that the histogram is additionally influenced by shell effects. We believe that this is not in contradiction to the analysis discussed here, because also in the case where electronic shell effects dominate, the nanocontact is composed of atoms and must accomodate the atomic structure. In addition, the histogram is generated from an ensemble of nanocontacts, where many individual atomic configurations add to the data. The interplay between shell effects and the atomic structure should be investigated further.

\section{Conclusion}

We have investigated Cu nanocontacts formed by electromigration thinning. We have identified peaks in conductance histograms resulting from the highly symmetric crystallographic directions. In addition we have analyzed the Fourier transform of the conductance histograms and found additional peaks due to hcp Cu and due to oxidation. Including hcp-Cu and Cu(1)oxide allow us to explain all peaks of the Fourier transform within the relevant frequency range. We check our results by using only part of the data. We view our efforts as in a way similar to powder diffractometry, where also the Fourier transform of an incident X-ray or electron beam is used, and the peaks are identified in terms of the structure of the object under study viewed in reciprocal space. By comparison with published data, we find hints for enhanced electromigration at defects occurring even at low current densities and low environment temperatures.

\section{Acknowledgements}
We thank Michael Marz and Christoph S\"urgers for help with the experimental work. This work was supported by the ERC Starting Grant NANOCONTACTS and by the ministry of science, research and arts, Baden-W\"urttemberg, in the framework of the Brigitte-Schlieben-Lange program.

\section{Appendix - Calculation of the periodicity of the conductance in hcp-structure and application on Cu}

\begin{figure}
	\includegraphics[width=0.6\linewidth]{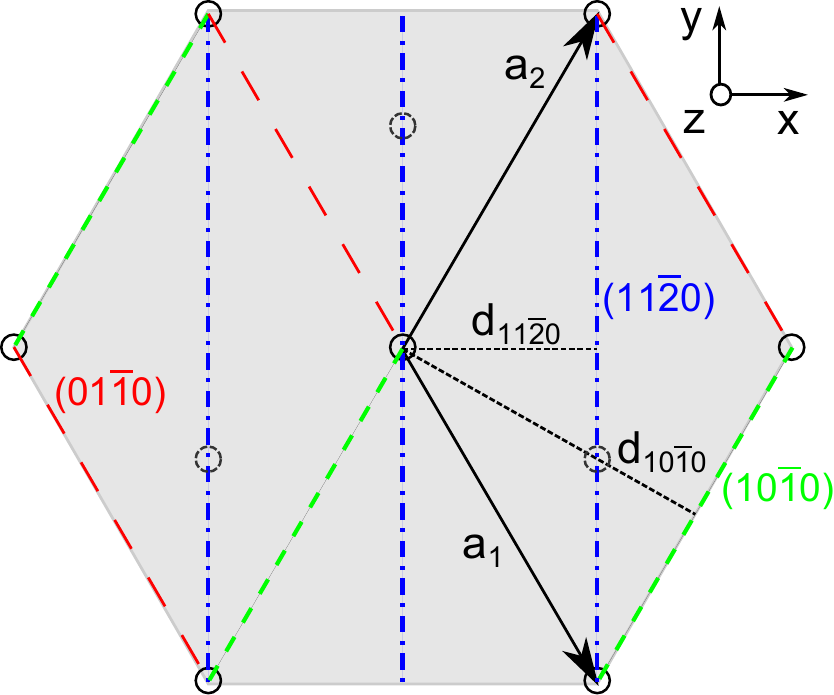}
	\caption{\label{fig-6} hcp-coordinates in the $\left( 0001 \right)$-plane. Atoms at the corner of the triangles are located in the basal plane and are marked as black circles. Atoms located in the center of one of the triangles are located at height $(1/2)c$ and are marked as dashed gray circles. The $\left(11\overline{2}0\right)$ planes are shown in blue, the $\left(10\overline{1}0\right)$ planes are marked in green.}
\end{figure}

The goal of this section is to calculate the periodicity $\Delta g_{ijkl}$ expected in the conductance histogram given by the Sharvin formula or variations of it \cite{Yanson2005} for a hexagonal lattice. This periodicity is related to the area an atom occupies in a certain crystallographic direction $\left( ijkl \right)$ and reads $\Delta g_{ijkl}=\pi \left(A_{ijkl} / \lambda_{\text{F}}^2\right)$. Here, we show how we calculate this area for the three highly symmetric lattice planes $\left( 10\overline{1}0 \right)$, $\left( 0001 \right)$ and $\left( 11\overline{2}0 \right)$. The hcp-lattice is a hexagonal lattice of edge length $a$ where each unit cell comprises two atoms, one at the corner of the hexagon on the basal plane and one in the center of one triangle at half of the unit cell height $c$. The area occupied by one atom in the plane $\left( ijkl \right)$ is given by the unit cell volume $v$ and the distance $d_{ijkl}$ between planes $\left( ijkl \right)$
\begin{equation} \label{eq:A1}
A_{ijkl}=\frac{v}{d_{ijkl}}
\end{equation}

\begin{figure}[t]
	\includegraphics[width=0.9\linewidth]{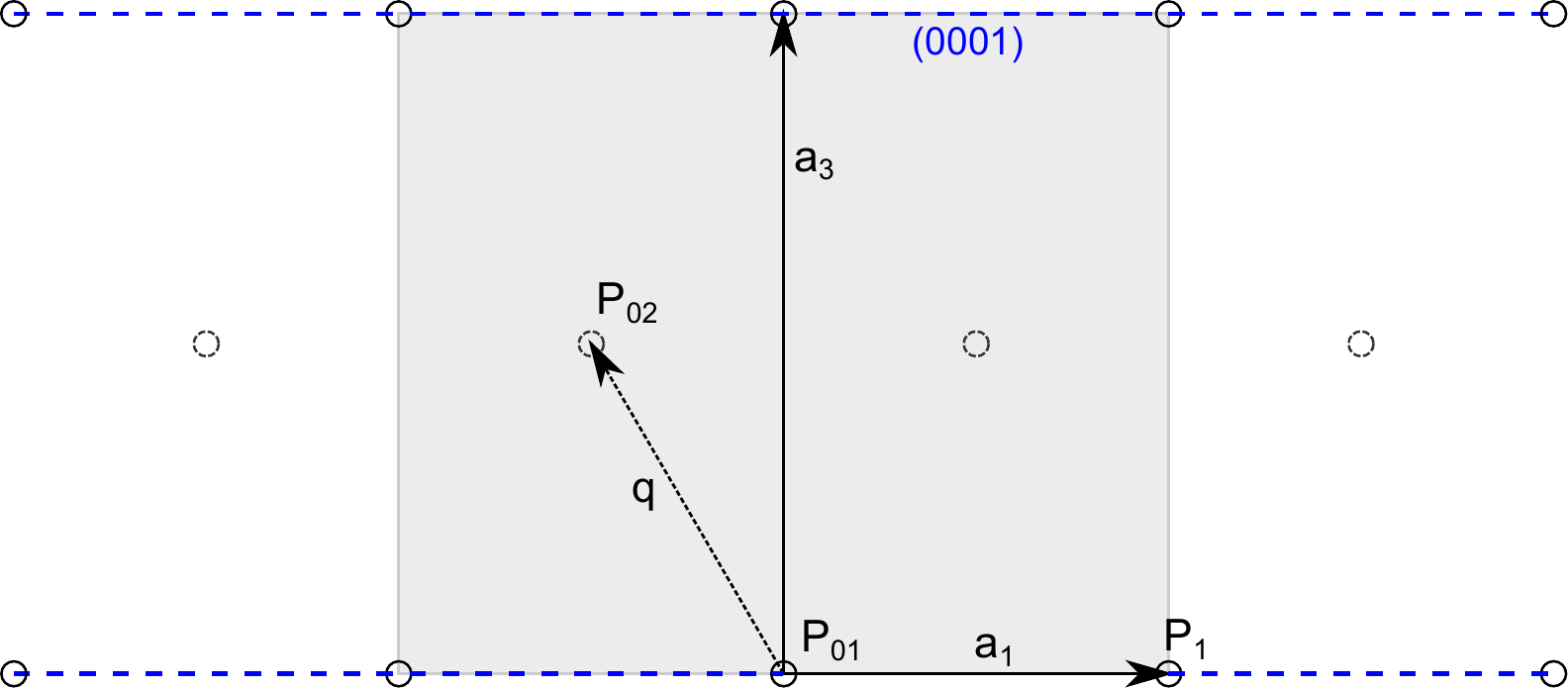}
	\caption{hcp-coordinates in the $\left( 01\overline{1}0 \right)$-plane. $\textbf{q}$ connects the two points $\text{P}_{01}$ und $\text{P}_{02}$ of the twoatomic basis. The two points in the middle are offset by $\frac{\sqrt{3}}{2}a$ out of the image plane.\label{fig-7}}
\end{figure}

The primitive vectors for a hexagonal lattice can be chosen as $\textbf{a}_1 = \left( \frac{a}{2},\, -\frac{\sqrt{3}}{2}a,\, 0 \right)^{\text{\textbf{T}}} $, $\textbf{a}_2 = \left(\frac{a}{2},\, \frac{\sqrt{3}}{2}a,\, 0 \right)^{\text{\textbf{T}}}$, see Fig. \ref{fig-6} and $\textbf{a}_3 = \left( 0,\, 0,\, c\right)^{\text{\textbf{T}}}$, see Fig. \ref{fig-7}. The unit cell volume $v$ is given by
\[
v = \textbf{a}_1 \cdot \left( \textbf{a}_2 \times \textbf{a}_3 \right) = \frac{\sqrt{3}}{2}a^{2}c
\]

The distance between the planes is easily determined geometrically (see figures \ref{fig-6} and \ref{fig-7}): $d_{10\overline{1}0} = \frac{\sqrt{3}}{2}a$, $d_{0001} = |\textbf{a}_3| = c$ and $d_{11\overline{2}0} =\frac{1}{2}a$.
To calculate the area occupied by an atom in the three planes we now use equation \ref{eq:A1}. For the planes $\left( 10\overline{1}0 \right)$ and $\left( 0001 \right)$ we use the full volume of the unit cell, even though the unit cell contains two atoms, because the planes only intersect with one of the two atoms. The plane $\left( 11\overline{2}0 \right)$ intersects with both atoms of the unit cell. For this plane the volume of half of the unit cell is used.

\[
A_{(10\overline{1}0)}=ac \text{, }\quad A_{(0001)} = \frac{\sqrt{3}}{2} a^{2} \quad\text{ and }\quad A_{(11\overline{2}1)} = \frac{\sqrt{3}}{2}ac
\]

For the periodicity follows
\begin{eqnarray}
\Delta g_{(10\overline{1}0)} & = & \frac{ack_{\text{F}}^2}{4\pi}\nonumber\\
\Delta g_{(0001)} & = & \frac{\sqrt{3}a^{2}k_{\text{F}}^{2}}{8\pi}\nonumber\\
\Delta g_{(11\overline{2}0)} & = & \frac{\sqrt{3}ack_{\text{F}}^2}{8\pi}\nonumber
\end{eqnarray}
With\cite{Jona2003}  $a_{\text{Cu;hcp}} = 2.956 \text{\,\AA}$, $c_{\text{Cu;hcp}} = 5.010 \text{\,\AA}$ and $k_{\text{F,Cu}} = 1.36 \text{\,\AA}^{-1}$ the expected periodicity of the conductance for hcp copper can now be determined.
\begin{subequations}
\label{eq:A2}
\begin{eqnarray}
\Delta g_{\text{hcp-Cu};(10\overline{1}0)} & \approx & 2.180 \label{eq:A2a} \\
\Delta g_{\text{hcp-Cu};(0001)} & \approx & 1.114 \label{eq:A2b} \\
\Delta g_{\text{hcp-Cu};(11\overline{2}0)} & \approx & 1.888  \label{eq:A2c}
\end{eqnarray}
\end{subequations}


\end{document}